\documentclass[twocolumn,showpacs,preprintnumbers,
amsmath,amssymb,aps,prd]{revtex4}
\def\be{\begin{equation}}
\def\ee{\end{equation}}
\def\bea{\begin{eqnarray}}
\def\eea{\end{eqnarray}}
\def\nn{\nonumber}
\def\p{\partial}
\def\half{\frac{1}{2}}
\def\mA{\mathcal{A}}

\begin{document}
\preprint{hep-th/0411002}
\title{The Conformal Version of Black Hole Thermodynamics}
\author{Shuang-Qing Wu$^*$} 
\affiliation{\centerline{Department of Physics, College of Physical Science and Technology,} \\
\centerline{Central China Normal University, Wuhan, Hubei 430079, People's Republic of China}}
\date{October 30, 2004}
\revised{\today}

\begin{abstract}
The conformal thermodynamics of rotating charged black holes in general relativity and string theory
is proposed by considering the first laws of thermodynamics for a pair of systems made up of the two
horizons of a Kerr-Newman or Kerr-Sen black hole. These two systems are constructed by only demanding
their ``horizon areas'' to be the sum and difference of that of the outer and inner horizons of their
prototype. The thermodynamics present here is a ``conformal version'' of black hole thermodynamics,
since it is closely related to the near-horizon conformal symmetry of black holes. The concept of
non-quasinormal modes recently proposed by D. Birmingham and S. Carlip \cite{BC2} is compatible with
this ``conformal thermodynamics'', rather than the usual ``horizon thermodynamics''. In addition, we
show that this conformal thermodynamics resembles to the thermodynamics of effective string or D-brane
models, since the two newly-constructed systems bear a striking resemblance to the right- and left-movers
in string theory and D-brane physics.
\end{abstract}

\pacs{04.70.Dy, 04.70.Bw, 11.25.-w}
\maketitle

It is well-known that black hole thermodynamics was established in the late 1960s and 1970s through a close
analogy between the ordinary laws of thermodynamics and certain laws of black hole mechanics \cite{BCH}. In
particular, the first law of black hole thermodynamics is formulated as a relationship between the first order
variation of the horizon area and those of global charges \cite{GC} such as the mass, angular momentum, and
electric charge. Since it correlates the thermodynamical quantities defined at the horizons of a black hole
with the global charges defined in the infinity, the formalism of the usual four laws of black hole mechanics
is a ``horizon version'' of black hole thermodynamics, also generally called as ``horizon thermodynamics''.
The first law makes it possible to assign an entropy to the horizon area and a temperature to the surface
gravity up to a coefficient \cite{BH}. Hawking's discovery \cite{SWH} of quantum particle creation by the
black hole event horizon put the first law of black hole thermodynamics on a solid fundament, and the
original analogy was seen to be of a physical nature.

The above ``horizon version'', however, has various limits on interpreting some black hole phenomena related
to the near-horizon symmetry of black holes, since it contains no thermodynamical quantity associated with this
symmetry. In the context of dual conformal field theory, Carlip \cite{SC} argued that black hole thermodynamics
and the statistical origin of entropy are controlled universally by the near-horizon conformal symmetry inherited
from the classical theory. This symmetry should play an important role in explaining the black hole phenomena
such as Hawking radiation \cite{SNS}, greybody factors \cite{GBF}, and more recently quasinormal modes \cite{BC1}
as well as non-quasinormal modes \cite{BC2}, etc. These dynamic processes cannot be well-understood in the
framework of the conventional ``horizon thermodynamics''. In particular, thermodynamical quantities related to the
near-horizon conformal symmetry of black holes rather than those defined on the horizons, enter into the greybody
factors, quasinormal modes and non-quasinormal modes. The quasinormal modes and non-quasinormal modes of a
three-dimensional BTZ black hole \cite{BTZ} can be interpreted in terms of the near-horizon conformal field
theory \cite{BC1, BC2}. It is reasonable to believe that the near-horizon geometry and near-horizon conformal
symmetry are crucial to understanding these properties of black holes.

On the other hand, both effective strings and D-brane models reveal that the Hawking temperature of the outer
horizon of a black hole is the harmonic mean of the effective temperatures of the right- and left-movers of
string \cite{CL}. In these theories, black holes can be modelled by effective or fundamental strings \cite{CL}
via representing them as bound states of D-branes that carry the same energy and charges as the black holes,
and the black hole entropy can be attributed to the sum of contributions from right- and left-moving excitations
of the string states. A characteristic feature of effective strings or D-branes is that their spectrum divides
into two distinct sectors associated with the right (R) and left (L) moving excitations, respectively. The
temperature of this combined system is introduced by hand as the harmonic mean of that of its parts:
\be\label{temp}
 \frac{2}{T_H} = \frac{1}{T_R} +\frac{1}{T_L} \, .
\ee
This temperature formula obviously demonstrates the importance of the near-horizon conformal symmetry, it holds
commonly in the dynamic phenomena such as absorption cross section, emission rates, decay rates, greybody factors
\cite{GBF}, and moreover quasinormal modes \cite{BC1} and non-quasinormal modes \cite{BC2} of black holes. Its
universality in all these cases cannot also be well-understood in the ``horizon version'' of black hole
thermodynamics.

In this paper, we propose a novel formalism of the first law of thermodynamics of rotating charged black holes.
In contrast to the usual ``horizon version'', the formulation present here is a ``conformal version'' of black
hole thermodynamics, since it is closely related to the near-horizon conformal symmetry of black holes. Our
motivation is also invoked by the recent interest on the very high damped quasinormal modes of Kerr black holes
\cite{QNM}. In a subsequent paper \cite{Wu}, we obtain the non-quasinormal modes of a Kerr-Newman black hole
analytically as functions of the right/left- temperatures, angular velocities, and electronic potentials (see Eq.
(\ref{nqnm}) below). We also show that the concept of non-quasinormal modes recently proposed in Ref. \cite{BC2}
is compatible with this ``conformal thermodynamics'', rather than the usual ``horizon thermodynamics'', when it
is applied to the case of rotating charged black holes \cite{Wu}. This result may help to determine the quasi-normal
modes of rotating black holes, since up till now there is no confirmable analytical result of Kerr black hole
quasi-normal modes.

To our aim, we take a new look at the classical thermodynamics of a Kerr-Newman (KN) \cite{KN} or Kerr-Sen
\cite{Sen} black hole and wish to seek some clues on relating it with string theory, especially with the
effective string/D-brane picture of black holes in a pure thermodynamic side. Specifically, we consider
the first law of thermodynamics of a pair of systems constructed from the two horizons of a KN black hole.
[Here we take a Kerr-Newman black hole as a typical representative. For a Kerr-Sen black hole, situation
is similar (see below).] It is interesting to observe that these two systems bear a striking resemblance
to the right- and left-movers in string theory and D-brane physics. We point out that there exists a close
relationship between the thermodynamics of these systems and that of effective string or D-brane models.
A remarkable advantage of our discussion presented here is that it relies on the validity in that the laws
of black hole mechanics does not depend upon the details of the underlying dynamical theory, even without
knowledge of fundamental string theory.
In the meanwhile, we also present a very general thermodynamical proof of the relations between the left/right-
temperatures and those of the inner/outer horizons (see Eq. (\ref{tem1}) below), while the same relation, namely
Eq. (\ref{temp}), is put by hand in the effective string/D-brane models.

To begin with, recall that the usual thermodynamics of a four-dimensional charged rotating (KN) black hole
with three classical parameters: the mass $M$, charge $Q$, and angular momentum $J = Ma$. For our discussion
here, we only focus on the first law of black hole thermodynamics in this most general case. It is essentially
formulated as the Bekenstein-Smarr (B-S) differential mass formula \cite{BS}:
\be
 dM = \half\kappa~d\mA +\Omega~dJ + \Phi~dQ \, ,
\ee
where $\kappa$, $\Omega$, and $\Phi$ denote, respectively, the surface gravity, angular velocity, and electrostatic
potential of the event horizon of a KN black hole. (Here we use the notation of a ``reduced'' horizon area $\mA =
A/(4\pi)$ rather than the horizon area $A$, so the Bekenstein-Hawking entropy is $S = A/4 = \pi\mA$.)

The KN black hole has two horizons: the inner Cauchy horizon $r_- = M -\epsilon$ and the outer event horizon
$r_+ = M +\epsilon$, where $\epsilon = \sqrt{M^2 -Q^2 -a^2}$.
In particular, the thermodynamics associated with the outer event horizon of the black hole is related to the
fundamental process of Hawking radiation. Similarly, one can prove that the thermodynamics of the inner Cauchy
horizon is associated with another quantum process of Hawking ``absorption'' \cite{WC,ZZL}. Namely, for an
observer rest at the infinity he observes a net flux of Hawking radiation outgoing from the event horizon
to the infinity, while for an imaginary observer inhabiting in the intrinsic singular region (inside the Cauchy
horizon), he will observe a flux of Hawking ``absorption'' ingoing from the intrinsic singular region to the
inner horizon also. Because a KN black hole is stationary, its outer horizon is in thermal equilibrium with
the thermal radiation outside the black hole. We think that its inner horizon is certainly in thermal
equilibrium with the thermal radiation in the intrinsic singular region \cite{ZZL}.

As is well-known, the outer horizon of a KN black hole is treated as a single thermodynamical system, and the
conventional four laws of thermodynamics are just stated for this system. Similarly, if one treats the inner
horizon as an independent thermodynamical system, one can establish another four laws of thermodynamics for
the inner horizon also \cite{WC}. The well-known B-S integral and differential mass formulae for the first
laws of thermodynamics corresponding to the inner and outer horizons of a general, non-extremal KN black hole
are given by \cite{WC,ZZL,CF}
\bea\label{FLio}
  M &=& ~~~~\kappa_{\pm}\mA_{\pm} +2\Omega_{\pm}J +\Phi_{\pm}Q \, , \nn \\
 dM &=& \half\kappa_{\pm}d\mA_{\pm} +\Omega_{\pm}dJ +\Phi_{\pm}dQ \, ,
\eea
where the ``reduced'' horizon area $\mA_{\pm} = r^2_{\pm} +a^2 = 2Mr_{\pm} -Q^2$, while the surface gravity
$\kappa_{\pm}$, angular velocity $\Omega_{\pm}$, and electric potential $\Phi_{\pm}$ at the two horizons are
\cite{WC}, respectively,
\be\label{KNQs}
 \kappa_{\pm} = \frac{r_{\pm} -M}{\mA_{\pm}} = \frac{\pm\epsilon}{\mA_{\pm}} \, , ~~
 \Omega_{\pm} = \frac{a}{\mA_{\pm}} \, , ~~ \Phi_{\pm} = \frac{Qr_{\pm}}{\mA_{\pm}} \, .
\ee
It is important to stress that the inner horizon is an abnormal negative temperature system due to $\kappa_- < 0$
\cite{WC,CF}, and moreover the inner horizon is much ``heater'' than the outer horizon as $|\kappa_-| > \kappa_+$
\cite{WC}.

In general, the B-S mass formulae can be derived from Christodoulou mass-squared formula \cite{CR}
for the reversible and irreversible processes by using thermodynamical relations. Specifically to a general,
nonextremal KN black hole, the Christodoulou formula \cite{CR} reads
\be\label{CR}
 M^2 = \frac{\mA_+ +\mA_-}{4} +\frac{Q^2}{2} = \frac{\mA_{\pm}}{4} +\frac{4J^2+Q^4}{4\mA_{\pm}} +\frac{Q^2}{2} \, .
\ee
From the identity (\ref{CR}), one can derive the expressions for the three thermodynamical quantities in Eq.
(\ref{KNQs}) by the following thermodynamical relations
\be
 \kappa_{\pm} = 2\frac{\p M}{\p\mA_{\pm}}\Big|_{J,Q}\, , ~
 \Omega_{\pm} = \frac{\p M}{\p J}\Big|_{Q,\mA_{\pm}}\, , ~
 \Phi_{\pm} = \frac{\p M}{\p Q}\Big|_{J,\mA_{\pm}}\, .
\ee

As a KN black hole is a system of two horizons, one should consider both of them for the sake of completeness.
Considerable research \cite{CL,ZZL,WC,CF} has confirmed this point. For instance, the importance of the inner
horizon is also emphasized in the viewpoint of string theory \cite{CL}. Just as did in Ref. \cite{ZZL}, we may
think a KN black hole can be regarded as a composite thermodynamical system composed of two independent subsystems,
its outer and inner horizons. For each subsystem, the entropy of each horizon is proportional to its horizon
area, and the temperature is proportional to its surface gravity, namely, $S_{\pm} = \pi \mA_{\pm}$, and
$T_{\pm} = \kappa_{\pm}/(2\pi)$.

Now we present our main idea. Using the original inner and outer horizons as blocks, we can construct two new
thermodynamical systems, called respectively as the R-system and L-system, by demanding their ``reduced'' areas
to be the sum and the difference of that of the outer and inner horizons of a KN black hole
\bea
 \mA_R &=& \mA_+ + \mA_- = 4M^2 -2Q^2 \, , \nn \\
 \mA_L &=& \mA_+ -\mA_- = 4M\epsilon \, . \quad
\eea
This idea is inspired by considerable researches done in Ref. \cite{CL} from the viewpoint of string theory,
where the left- and right-moving thermodynamics of the string theory correspond to the sum and the difference
of the outer and the inner horizon thermodynamics. However it is reverse to the observation of that of Ref.
\cite{CL}. We mention that our idea is not only supported from studies on quantum mechanics and area spectrum
of black holes \cite{MR, VW}, but also invoked by the attempt to re-define a new black hole entropy (corresponding
to the entropy of L-system here) which satisfies the Nernst theorem \cite{ZZL}.

We would like to see whether or not the first law of thermodynamics of the original KN black hole can be understood
through exploring the thermodynamic features of these two new systems. In one way each system can be identified with
a black hole. As the original motherboard carries three classical hairs ($M, Q, J$), by the definition of our systems,
it is immediately recognized that the R-system carries only two hairs ($M, Q$), and the L-system carries three hairs
($M, Q, J$). So they are two distinct copies describing two different kinds of black holes. In particular, there
exists an asymmetry of angular momentum between the R- and L-systems as we shall see below. The disappearance of
angular momentum in the R-system and its presence in the L-system imply that the black hole degrees of freedom
relevant for near-BPS excitations can be described by a $(0, 4)$ chiral superconformal field theory with an SU(2)
current algebra associated to rotations \cite{GBF}.

It is now a position to reveal the thermodynamic properties of our new systems. To investigate their thermodynamic
features in detail, let us focus firstly on the R-system. For this system, the Christodoulou-type mass-squared
formula \cite{CR} is simply written as
\be\label{CR1}
 M^2 = \frac{\mA_R}{4} +\frac{Q^2}{2} \, .
\ee
From this relation, it is easily to deduce the integral and differential B-S mass formula as follows:
\bea\label{FL1}
  M &=& \frac{\mA_R}{4M} +\frac{Q^2}{2M} ~= ~~~~\kappa_R\mA_R +\Phi_R Q \, , \nn \\
 dM &=& \frac{d\mA_R}{8M} +\frac{Q dQ}{2M} = \half\kappa_R d\mA_R +\Phi_R dQ \, ,
\eea
where the surface gravity $\kappa_R = 1/(4M)$, and electric potential $\Phi_R = Q/(2M)$ can also be obtained
by the following thermodynamical definitions
\bea
 && \kappa_R  = 2\frac{\p M}{\p\mA_R}\Big|_Q = \frac{1}{4M} = \frac{\epsilon}{\mA_L} \, , \nn \\
 && \Phi_R  = \frac{\p M}{\p Q}\Big|_{\mA_R} = \frac{Q}{2M} = \frac{2Q\epsilon}{\mA_L} \, .
\eea

The R-system can be viewed as describing a nonrotating charged black hole with mass $M$ and charge $Q$, while
its entropy and temperature being given by
\bea\label{ent1}
 S_R &=& \pi\mA_R = \pi(\mA_+ +\mA_-) = S_+ +S_- \nn \\
     &=& 2\pi(2M^2 -Q^2) \, , \nn \\
 T_R &=& \frac{\kappa_R}{2\pi} = \frac{1}{8\pi M} \, .
\eea
It is not difficult to recognize that this system is thermodynamically equivalent to the GHS dilatonic black
hole \cite{GHS} with the dilaton parameter $\phi_0 = 0$, whose horizon is at $r_h = 2M$, its surface gravity
is $\kappa_h = 1/(4M)$, and ``reduced'' horizon area is just $\mA_h = r_h(r_h -Q^2/M) = \mA_R$.

Now we turn to the L-system. The above derivation is also applicable to this case, but with a little involved
algebras. The corresponding Christodoulou-type formula \cite{CR} is given by
\be\label{CR2}
 M^2 = \frac{Q^2 +\sqrt{Q^4 +4J^2 +\mA_L^2/4}}{2} \, ,
\ee
and the first law is formulated as follows
\bea\label{FL2}
  M &=& ~~~~\kappa_L\mA_L +2\Omega_L J +\Phi_L Q \, , \nn \\
 dM &=& \half\kappa_L d\mA_L +\Omega_L dJ +\Phi_L dQ \, ,
\eea
where the surface gravity $\kappa_L$, angular velocity $\Omega_L$, and electric potential $\Phi_L$ for this
system are determined by the following thermodynamical expressions
\bea
 && \kappa_L = 2\frac{\p M}{\p\mA_L}\Big|_{J,Q} = \frac{\epsilon}{\mA_R}
  = \frac{\sqrt{M^2 -Q^2 -a^2}}{4M^2 -2Q^2} \, , \nn \\
 && \Omega_L = \frac{\p M}{\p J}\Big|_{Q,\mA_L} = \frac{2a}{\mA_R} = \frac{a}{2M^2 -Q^2} \, , \\
 && \Phi_L = \frac{\p M}{\p Q}\Big|_{J,\mA_L} = \frac{2QM}{\mA_R} = \frac{QM}{2M^2 -Q^2} \, . \nn
\eea
To check that the above expressions hold true, an identity $\mA_+\mA_- = (\mA_R^2 -\mA_L^2)/4 = Q^4 +4J^2$
might be used.

The L-system can represent a rotating charged black hole with mass $M$, charge $Q$, and angular momentum $J$,
its entropy and temperature are given by
\bea\label{ent2}
 S_L &=& \pi\mA_L = \pi(\mA_+ -\mA_-) = S_+ -S_- \nn \\
     &=& 4\pi M\sqrt{M^2 -Q^2 -a^2} \, , \nn \\
 T_L &=& \frac{\kappa_L}{2\pi} = \frac{\sqrt{M^2 -Q^2 -a^2}}{4\pi(2M^2 -Q^2)} \, .
\eea
An important feature of the L-system is that it can reflect the extremality of a KN black hole when it
approaches to its extremal limit, namely, the extremal KN black hole.

It is remarkable to seek the relations between the temperatures of the two newly-constructed systems and
those of the two horizons of the original KN black hole. It is easy to obtain the following relations
\bea\label{tem1}
&& \frac{1}{\kappa_R} = \frac{\mA_L}{\epsilon} = \frac{\mA_+ +\mA_-}{\epsilon}
  = \frac{1}{\kappa_+} +\frac{1}{\kappa_-} \, , \nn \\
&& \frac{1}{\kappa_L} = \frac{\mA_R}{\epsilon} = \frac{\mA_+ -\mA_-}{\epsilon}
  = \frac{1}{\kappa_+} -\frac{1}{\kappa_-} \, ,
\eea
which have an \textit{exact} correspondence in string theory and D-brane physics \cite{GBF}. In particular,
$\kappa_+$ bears the same harmonic mean relation as $T_H$ in Eq. (\ref{temp}). Here we present a very general
thermodynamical proof of the relations between the left/right- temperatures and those of the inner/outer
horizons, while the same relation (\ref{temp}) is put by hand in the effective string/D-brane models.
In addition, we find that the angular velocities and electric potentials between the new and old systems
are related by
\bea
 && \frac{2}{\Omega_L} = \frac{\mA_R}{a} = \frac{\mA_+ +\mA_-}{a}
  = \frac{1}{\Omega_+} + \frac{1}{\Omega_-} \, , \nn \\
 && \Phi_R = \frac{\Phi_+ +\Phi_-}{2} +\Big(\frac{\kappa_R}{\kappa_L}\Big)
  \frac{\Phi_+ -\Phi_-}{2} \, , \\
 && \Phi_L = \frac{\Phi_+ +\Phi_-}{2} +\Big(\frac{\kappa_L}{\kappa_R}\Big)
  \frac{\Phi_+ -\Phi_-}{2} \, . \nn
\eea

In the Schwarzschild black hole case, the R-system and L-system are indistinguishable, both of them are
identical to its prototype. In other words, the two systems are in thermal equilibrium. It should be
stressed that the above reformulation of the first law of classical black hole thermodynamics are universal
for large classes of black holes in various dimensions. For a Kerr-Sen black hole arising in the low energy
effective heterotic string theory, one can obtain similar results with some slight modifications. [$\mA_{\pm}
= 2Mr_{\pm} = 2M(M -b \pm \varepsilon)$, $\Phi_{\pm} = \Phi_{R,L} = Q/(2M) = b/Q$, where $\varepsilon =
\sqrt{(M-b)^2 -a^2}$ in place of $\epsilon$.] Similar formulation can be established in the case of a
three-dimensional BTZ black hole \cite{BTZ} and in other dimensions also.

Note that there is an asymmetry of angular momentum between the R- and L-systems. This fact reflects that
unlike the 5-dimensional case, the addition of angular momentum in four dimension always breaks the right-moving
supersymmetry \cite{GBF}.
Without rotation, the D-brane description of an extremal Reissner-Nordstr\"{o}m black hole can be modelled
by a $1+1$-dimensional field theory which turns out to be a $(4, 4)$ superconformal sigma model. When the
angular momentum is included, a black hole can be extremal, but supersymmetry is broken, leaving us with
$(0, 4)$ superconformal symmetry \cite{GBF}. The $N = 4$ superconformal algebra gives rise to a left-moving
$SU(2)$ symmetry of the original $(4, 4)$ chiral superconformal field theory.
The $(0, 4)$ chiral supersymmetry algebra contains a left-moving $SU(2)_R$ symmetry, which in fact is the same
as the $SU(2)$ symmetry of global spatial rotations, a symmetry of the Hilbert space of black hole states. The
charge under one $U(1)$ subgroup of this $SU(2)$ group will then be related to the four-dimensional angular
momentum carried by the left movers. The right movers, however, cannot carry macroscopic angular momentum.

The physical meaning or consequences of the thermodynamics of these two new systems is as follows. In the
light of R-system, a Reissner-Nordstr\"{o}m, a Kerr-Newman, a Kerr-Sen, and a GHS dilatonic black hole are
thermodynamically equivalent to each other, (a Kerr black hole is equivalent to a Schwarzschild black hole);
while they are distinct in the light of L-system. Besides, the temperature and area of our R-system are
consistent with those obtained in Ref. \cite{MR}; the temperature and area of Ref. \cite{VW} coincide with
that of our L-system.

In addition, it may provide a new routine to understand thermodynamics of extremal black holes. Although we
deal with the general, non-extremal Kerr-Newman black holes, extremal and near-extremal Kerr-Newman black
holes are also included here as special cases. In the extremal Kerr-Newman black hole case, according to the
usual ``horizon thermodynamics'', it has zero temperature $T_e = 0$ and nonzero horizon area $\mA_e = 2M^2 -Q^2$;
but according to the thermodynamics presented here, it has nonzero temperature $T_R = 1/(8\pi M)$ and nonzero
area $\mA_R = 4M^2 -2Q^2$ in the R-system, while it has zero temperature $T_L = 0$ and zero area $\mA_L = 0$
in the L-system. So our theory opens a new eye on the controversy about the entropy of an extremal black hole,
namely, whether its entropy is zero or nonzero.

Distinguished from the ``horizon version'' of black hole thermodynamics, the Bekenstein-Smarr formulae (Eqs.
(\ref{FL1}) and (\ref{FL2})) relate the same global charges but with the thermodynamical quantities related
to the conformal properties of near-horizon geometry. So the thermodynamics developed here is a ``conformal
version'' of black hole thermodynamics since it is closely related to the near-horizon conformal symmetry
of black holes. It may be called as ``conformal thermodynamics''. Dynamic phenomena mentioned above
\cite{SNS,GBF,BC1,BC2} can be easily understood in the language of this ``conformal thermodynamics''. In a
subsequent paper \cite{Wu}, we show that the concept of non-quasinormal modes recently proposed in \cite{BC2}
is compatible with this ``conformal thermodynamics'', rather than the usual ``horizon thermodynamics''. By
calculating the monodromies of inner/outer horizons, the non-quasinormal modes of a Kerr-Newman black hole
are analytically obtained as \cite{Wu}
\bea\label{nqnm}
&& \omega_R = m\Omega_R +q\Phi_R -2i\kappa_Rn \, , \nn \\
&& \omega_L = m\Omega_L +q\Phi_L -2i\kappa_Ln \, ,
\eea
where $m$ is the azimuthal quantum number, $q$ is the charge of external field, and $n$ is a positive integer.
This new result may be helpful to determine the quasi-normal modes of rotating black holes, since there is
no confirmable analytical result of a Kerr black hole up till now. The analytical formula conjectured in Ref.
\cite{SHod} is inconsistent with new numerical results \cite{QNM}. The recent new numerical results show that
the real and imaginary parts of very high damped quasi-normal modes are not a simple function of the temperature
and angular velocity of the outer horizon \cite{QNM}. Our result will provide a very useful guidance to this
important question.

Next we would like to point out that the above R- and L-systems bear a striking resemblance to the right-
and left-movers in string theory and there exists a close relationship between the thermodynamics of these
systems and that of effective strings or D-branes. Our reformulation of the first law of black hole
thermodynamics can be viewed as an analogy of thermodynamics of effective string or D-brane models.

Consider now the thermodynamics of effective strings or D-branes. The thermodynamical object often studied
in string theory and D-brane physics is an effective $1+1$-dimensional ideal gas system composed of a pair
of weakly-coupled oppositely moving states \cite{GMAS}. [In the cases of D-branes, the R- and L-moving
massless open string states on a long brane constitute two noninteracting one-dimensional gas of strings.]
Note that a black hole is in effect a thermodynamical system with an effective dimension $d = 1$ \cite{GM}.
The effective spatial dimension is one in both cases where the combined system of effective string is
divided into two independent or weak coupling R- and L-sectors. Each sector is a collection of excitations
modes and may be attributed to a gas of strings. The entropy of the combined system is interpreted as a
sum of two independent microscopic contributions from R- and L-moving modes, respectively. Since the
thermal equilibrium is assumed to maintain in each sector independently, so each sector can be endowed
with an effective temperature. Massless particles in one spatial dimension are either R-moving or L-moving,
all thermodynamical quantities can be split into a R-moving and a L-moving piece: $E = E_R + E_L$, etc.
For simplicity, we only consider the energy and entropy of two sectors and drop other thermodynamical
quantities (for example, by taking a vanishing chemical potential). It follows that the first laws for
the R- and L-sectors take form
\be
 dE_R = T_R dS_R \, , \qquad dE_L = T_L dS_L \, .
\ee
Taking their sum and difference, we obtain

\be\label{eas}
 dS_{\pm} = dS_R \pm dS_L = \frac{1}{T_{\pm}}dE_+ +\frac{1}{T_{\mp}}dE_- \, ,
\ee
where we introduce $E_{\pm} = E_R \pm E_L$, $S_{\pm} = S_R \pm S_L$, and
\be\label{tem2}
 \frac{2}{T_+} = \frac{1}{T_R} +\frac{1}{T_L} \, , \qquad
 \frac{2}{T_-} = \frac{1}{T_R} -\frac{1}{T_L} \, .
\ee

We further assume each sector carries energy $E_R = E_L = M/2$, then the total energy is $E = E_+ = M$,
and the total momentum is conserved: $P = -E_- = 0$. Thus Eq. (\ref{eas}) reduces to $dE = dE_+ = T_{\pm}
dS_{\pm}$. By comparing it with the formula $dM = \half\kappa_{\pm} d\mA_{\pm} +\cdots$ in Eq. (\ref{FLio})
and observing that Eq. (\ref{tem2}) is essentially the same one as in Eq. (\ref{tem1}), we establish that
\be
 T_{\pm} = \frac{\kappa_{\pm}}{2\pi} \, , \qquad
 S_{\pm} = \pi\mA_{\pm} = \frac{A_{\pm}}{4} \, .
\ee

We find that if the effective strings or D-branes carry the same energy and charges as the black hole, then
their right- and left-moving thermodynamics corresponds \textit{exactly} to the sum and difference of the outer
and inner horizons thermodynamics. This hints that the effective string or D-brane picture of black hole mechanics
may be universal in all cases where the correspondence between black holes and effective strings/D-branes
has been demonstrated. Indeed one may identify the constructed R- and L-systems with the R- and L-movers
in string theory. In this way it is natural to appreciate that black hole entropy can be explained within
string theory and D-brane physics by prescribing a microscopic description for all black holes.

In summary, we have proposed a new reformulation of classical black hole thermodynamics, which bears
a close analogy with ``stringy'' thermodynamics. The newly-made systems resemble to the R- and L-movers in
string theory and D-brane physics very much. In our opinion, this resemblance seems to be of a true nature,
so it provides a further evidence to support the effective string/D-brane picture of black holes in the pure
classical thermodynamic side. The formalism presented here suggests a plausible thermodynamic routine to
understand why the mysterious origin problem of black hole entropy could be resolved in the framework of
string theory and D-brane physics. In addition, it may setup a bridge to synthesize various related issues
such as the near-horizon conformal symmetry \cite{SC}, geometric ``conformal field theory'' \cite{CL},
quasinormal and non-quasinormal modes \cite{BC1,BC2,QNM}, and black hole entropy. Other important applications
might include area spectrum or area quantization \cite{MR, VW}, quasi-normal modes and thermodynamics of a
black hole with multi-horizons or degenerate horizon \cite{MdH}. It might also be associated with thermodynamics
of dynamical horizon \cite{DH}. We hope in the next step to reveal the explicit relation between the near-horizon
conformal symmetry and the conformal thermodynamics developed here.

This project was supported in part by China Postdoctoral Science Foundation under Grant No. 2002032234 and K. C.
Wong Education Foundation, Hong Kong.

\end{document}